\documentclass[a4paper,USenglish, autoref, thm-restate]{lipics-v2021}

\pdfoutput=1 %
\hideLIPIcs  %

\usepackage{lucas}

\usepackage{csvsimple} 
\usepackage[textsize=tiny]{todonotes}
\usepackage{csquotes}
\usepackage{nicematrix}
\usepackage[compat=0.6]{yquant}
\usepackage[binary-units]{siunitx}
\useyquantlanguage{groups}
\usepackage{booktabs}

\newtheorem{myexample}[theorem]{Example}

\title{Towards a SAT Encoding for Quantum Circuits}
\subtitle{A Journey From Classical Circuits to Clifford Circuits and Beyond}
\titlerunning{Towards a SAT Encoding for Quantum Circuits} 

\author{Lucas Berent}{Technical University of Munich, Germany}{lucas.berent@tum.de}{https://orcid.org/0000-0002-2973-1689}{}
\author{Lukas Burgholzer}{Johannes Kepler University Linz, Austria}{lukas.burgholzer@jku.at}{https://orcid.org/0000-0003-4699-1316}{}
\author{Robert Wille}{Technical University of Munich, Germany \and Software Competence Center Hagenberg GmbH (SCCH), Austria}{robert.wille@tum.de}{https://orcid.org/0000-0002-4993-7860}{}

\authorrunning{L. Berent, L. Burgholzer, and R. Wille} %

\Copyright{Lucas Berent, Lukas Burgholzer, and Robert Wille} %



\supplement{\emph{Source Code:} \href{https://github.com/lucasberent/qsatencoder}{github.com/lucasberent/qsatencoder}}%

 \funding{This work received funding from the European Research Council (ERC) under the European Union’s Horizon 2020 research and innovation program (grant agreement No. $101001318$), was part of the Munich Quantum Valley, which is supported by the Bavarian state government with funds from the Hightech Agenda Bayern Plus, and has been supported by the BMK, BMDW, and the State of Upper Austria in the frame of the COMET program (managed by the FFG).}

\nolinenumbers %

\begin{document}
\usetikzlibrary{quotes}
\maketitle

\begin{abstract}
\emph{Satisfiability Testing} (SAT) techniques are well-established in classical computing where they are used to solve a broad variety of problems, e.g., in the design of classical circuits and systems.
Analogous to the classical realm, quantum algorithms are usually modelled as circuits and similar design tasks need to be tackled. 
Thus, it is natural to pose the question whether these design tasks in the quantum realm can also be approached using SAT techniques.
To the best of our knowledge, no SAT formulation for arbitrary quantum circuits exists and it is unknown whether such an approach is feasible at all. 
In this work, we define a propositional SAT encoding that, in principle, can be applied to arbitrary quantum circuits. 
However, we show that due to the inherent complexity of representing quantum states, constructing such an encoding is not feasible in general. 
Therefore, we establish general criteria for determining the feasibility of the proposed encoding and identify classes of quantum circuits fulfilling these criteria. We explicitly demonstrate how the proposed encoding can be applied to the class of Clifford circuits as a representative.
Finally, we empirically demonstrate the applicability and efficiency of the proposed encoding for Clifford circuits. %
With these results, we lay the foundation for continuing the ongoing success of SAT in classical circuit and systems design for quantum circuits.
\end{abstract}

\section{Introduction}\label{sec:intro}
Quantum computing~\cite{nielsenQuantumComputationQuantum2010} has recently gained far-reaching interest in academia and industry due to the potential advantage in efficiency compared to classical computers for many practically relevant problems. 
There are several important problems in computer science and related fields such as physics, chemistry, and mathematics, for which it is known that quantum algorithms offer significant improvements over classical algorithms~\cite{bravyi2020quantum,groverFastQuantumMechanical1996,huang2021information,riste2017demonstration, shorPolynomialtimeAlgorithmsPrime1997}. 
In order to efficiently and correctly realize such quantum algorithms on actual quantum computers, a multitude of circuit design tasks needs to be addressed accordingly. Central problems in circuit design are, amongst others, compilation~\cite{amyStaqFullstackQuantum2020,hanerSoftwareMethodologyCompiling2018,smithQuantumComputationalCompiler2019}, synthesis~\cite{degriendArchitectureawareSynthesisPhase2020,gilesExactSynthesisMultiqubit2013,niemannEfficientSynthesisQuantum2014}, technology mapping~\cite{liTacklingQubitMapping2019,tanOptimalLayoutSynthesis2020,willeMappingQuantumCircuits2019}, simulation~\cite{brennanTensorNetworkCircuit2021,jaquesLeveragingStateSparsity2021,zulehnerAdvancedSimulationQuantum2019},  and verification~\cite{burgholzerAdvancedEquivalenceChecking2021,wangXQDDbasedVerificationMethod2008, yamashitaFastEquivalencecheckingQuantum2010}. Many of these problems have high worst-case complexity---some have even been proven to be NP-complete~\cite{boteaComplexityQuantumCircuit2018} or QMA-complete\footnote{The complexity class QMA is the quantum analogue to the classical complexity class NP~\cite{bookatzQMAcompleteProblems2013}.}~\cite{janzingNonidentityCheckQMAcomplete2005}. Hence, efficient methods to tackle practically relevant instances are needed.

In the classical realm, solvers for \emph{Boolean Satisfiability Testing}~(SAT,~\cite{biereHandbookSatisfiability2009}) are one of the key means to efficiently solve design tasks for the realization of classical circuits~\cite{biereSATBasedModelChecking2018,brandVerificationLargeSynthesized1993,eggersglussImprovedSATbasedATPG2013,DBLP:conf/date/GebregiorgisT19,DBLP:conf/fmcad/KaufmannBK19,larrabeeTestPatternGeneration1992,willeSMTbasedStimuliGeneration2009}. 
All of these SAT-based approaches rely on symbolically encoding the functionality of a given (classical) circuit into a propositional formula which (enriched by further constraints encoding the design objective) is passed to a SAT solver. 
Because of the remarkable improvements of SAT solvers in the past decades, modern solvers are able to efficiently reason over large formulas and therefore compute the desired design solutions. %

Having this power of efficient logical reasoning at hand, it is natural to wonder whether the prospects of \mbox{SAT-based} solutions can also be materialized for the complex design tasks outlined above for quantum circuits.
However, while encoding a classical circuit is rather straight-forward (each signal is represented by a propositional variable; gates are symbolically encoded through corresponding propositional formulas), quantum circuits rely on so-called qubits that do not only assume discrete values $0$ and $1$, but also \emph{superpositions} (i.e., \mbox{complex-valued} linear combinations) of both---creating an infinitely large state space even for a single qubit. 
Quantum-mechanical phenomena such as entanglement~\cite{nielsenQuantumComputationQuantum2010} further complicate the representation of states.
This raises the question whether a generalization of SAT encodings for quantum circuits is possible and, furthermore, if there exist SAT-based approaches similar to those that have become one of the most established techniques in the classical realm\footnote{In the domain of quantum computing, first SAT-based approaches tackling a particular combinatorial problem have successfully been proposed (e.g., in~\cite{tanOptimalLayoutSynthesis2020,willeMappingQuantumCircuits2019}). 
To the best of our knowledge, no complete SAT encoding for the functionality of quantum circuits exists.
Current techniques are either limited to reversible circuits~\cite{willeATPGReversibleCircuits2011,yamashitaFastEquivalencecheckingQuantum2010} (an important subclass of quantum circuits) or quantum circuits prohibiting entanglement~\cite{willeCompactEfficientSAT2013} (one of the core traits of quantum computing).}.

In this work, we tackle these questions. More precisely, we
\begin{itemize}
	\item provide a problem analysis and discussion on the limitations of (\mbox{straight-forward}) adaptations of classical SAT techniques to encode the functionality of quantum circuits,
	\item propose a new \emph{generalized encoding} that can be used to encode arbitrary quantum circuits and show that this increased capability comes at a certain price due to the sheer complexity of representing quantum states,
	\item identify classes of quantum circuits for which the proposed generalized encoding can be constructed efficiently, and %
	\item provide an empirical analysis on the scalability of the proposed satisfiability encoding and demonstrate its feasibility through experimental evaluations based on a proof-of-concept implementation.
\end{itemize}
With this work, we lay the foundation for further research towards leveraging powerful classical SAT techniques for quantum computing. 
In contrast to the classical realm and due to the inherent complexity of quantum states and operations, our work indicates that dedicated 
classes of quantum circuits will need to be considered in order to formulate efficient and scalable SAT encodings for design tasks involving quantum circuits.

The remainder of this work is organized as follows: 
\autoref{sec:background} introduces the necessary background to keep this work self-contained.
Then, \autoref{sec:motiveAndNaive} reviews how a SAT encoding for classical circuits is derived and how it can be adapted to (certain) quantum circuits.
Based on that,~\autoref{sec:new_endoding} describes a generalized SAT encoding and discusses its limitations.
Afterwards, \autoref{sec:cliffordAndBeyond} shows how to overcome these limitations for certain classes of quantum circuits.
\autoref{sec:experiments} summarizes our empirical analysis, before \autoref{sec:conclusion} concludes the paper.

\section{Background}\label{sec:background}
In this section, we briefly review the main concepts of quantum computing needed throughout the rest of this work. While the individual descriptions are kept brief, we refer the interested reader to~\cite{nielsenQuantumComputationQuantum2010} for a detailed introduction.

In classical computing, the fundamental unit of information is a bit, which can assume any of the Boolean values $0$ or $1$.
The analogue in the quantum realm is called quantum bit (or \emph{qubit}), which cannot only assume any of the computational basis states $\ket{0}$ or $\ket{1}$ but also arbitrary complex-valued linear combinations (\emph{superposition}) of these states.
More specifically, the state~$\ket{\varphi}$ of a single qubit is described as $\ket{\varphi} = \alpha_0 \ket{0} + \alpha_1 \ket{1}$ with $\alpha_0, \alpha_1\in\mathbb{C}$ and $\abs{\alpha_0}^2 + \abs{\alpha_1}^2 = 1$.
The complex-valued factors $\alpha_i$ are called \emph{amplitudes} and it is convenient to represent the state of a quantum system by a vector of amplitudes (the \emph{state vector}), i.e., $\ket{\varphi} \equiv \begin{bmatrix}\alpha_0 &\alpha_1\end{bmatrix}^\top$.
The postulates of quantum mechanics state that the state vector cannot be observed directly.
Instead, \emph{measuring} a qubit collapses its state to one of the (classical) basis states $\ket{i}$---each with probability $\abs{\alpha_i}^2$.

\begin{myexample}
    Consider the plus and minus states $\ket{+}$ and $\ket{-}$ which are represented by the state vectors $1/\sqrt{2}\begin{bmatrix}1&1\end{bmatrix}^\top$ and $1/\sqrt{2}\begin{bmatrix}1&-1\end{bmatrix}^\top$, respectively.
    Both states describe an equal superposition of the computational basis states. Measuring them yields either $\ket{0}$ or $\ket{1}$---each with a probability of $\abs{\pm 1/\sqrt{2}}^2 = 0.5$.
\end{myexample}

The basis states of an $n$-qubit system are formed by the tensor product (denoted $\otimes$ in the following) of single-qubit states, i.e., $\ket{i_{n-1}} \otimes \dots \otimes \ket{i_0} \equiv \ket{i_{n-1}\dots i_0} \equiv \ket{\sum_{j=0}^{n-1} 2^j i_j} \equiv \ket{i}$ with $i_{n-1},\dots,i_0\in\{0,1\}$ and $i\in\{0,\dots,2^n-1\}$.
Any $n$-qubit state $\ket{\varphi}$ is then described as an arbitrary superposition of these basis states, i.e., $\ket{\varphi} = \sum_{i=0}^{2^n-1} \alpha_i \ket{i}$ with $\alpha_i\in\mathbb{C}$ and $\sum_{i=0}^{2^n-1} \abs{\alpha_i}^2 = 1$.
Again, this is conveniently represented by the state vector, i.e., $\ket{\varphi}\equiv\begin{bmatrix}\alpha_{0\dots 0} & \dots & \alpha_{1 \dots 1}\end{bmatrix}^\top$. 
One of the most fundamental differences of quantum states to classical states is that the individual qubits of a system can be \emph{entangled}, i.e., their state can no longer be considered separately (as e.g., for computational basis states), but has to be considered as a whole.

\begin{myexample}\label{ex:bellstates}
The four Bell states $\ket{\Phi^\pm} = 1/\sqrt{2}(\ket{00} \pm \ket{11})$ and $\ket{\Psi^\pm} = 1/\sqrt{2}(\ket{01} \pm \ket{10})$ are one of the most prominent examples of entangled quantum states.
Consider, for example, the $\ket{\Phi^+}$ state and assume that the first of its qubits is measured.
The measurement collapses the state and leaves the system either in the state $\ket{00}$ or $\ket{11}$---each with a probability of $\abs{\pm 1/\sqrt{2}}^2 = 0.5$.
Consequently, the state of the second qubit is completely determined by the measurement result of the first qubit---without ever being \enquote{touched}.
\end{myexample}

Similarly to how classical operations and logic gates are applied to the bits of a classical system, \emph{quantum operations} or \emph{quantum gates} can be used to change the state of a quantum system.
To this end, a quantum operation acting on $k$ qubits is described by a complex-valued unitary\footnote{A matrix $U\in\mathbb{C}^{2^k\times 2^k}$ is \emph{unitary} if $U^{\dagger} U = UU^{\dagger} = I$, where $U^{\dagger}$ is the complex-conjugate of $U$ and $I$ denotes the identity matrix.} matrix $U\in\mathbb{C}^{2^k\times 2^k}$. 

\begin{myexample}
    One of the most fundamental single-qubit quantum operations are the Pauli gates $X$, $Y$, and $Z$, the phase gate $S$, as well as the Hadamard gate $H$, which are described by
    $$
    X=\begin{bmatrix}0&1\\1&0\end{bmatrix}, \; Y=\begin{bmatrix}0&-i\\i&0\end{bmatrix}, \; Z=\begin{bmatrix}1&0\\0&-1\end{bmatrix}, \; S=\begin{bmatrix}1&0\\0&i\end{bmatrix}, \; H=\frac{1}{\sqrt{2}}\begin{bmatrix}1 & 1\\1 & -1\end{bmatrix}.
    $$
    Some useful identities are $Z=SS$, $X=HZH$, and $Y=iXZ$. 
    An important example of a two-qubit operation is the controlled-NOT or $\mathit{CNOT}$ operation, which flips the state of a designated target qubit if the designated control qubit is in state $\ket{1}$. We write $\mathit{CNOT}_{c,t}$ for a CNOT gate controlled on qubit $q_c$ and targeted at qubit $q_t$. The corresponding $4\times4$ matrix is given by $\mathit{diag}(I, X)$.
\end{myexample}
Applying a quantum operation to the quantum state~$\ket{\varphi}$ corresponds to the \mbox{matrix-vector} product of the respective matrix~$U$ with the state vector representing $\ket{\varphi}$---yielding a new quantum state~$\ket{\varphi'}$\footnote{Technically, for the multiplication to make sense, any operation acting only on $k < n$ qubits must be extended to the full system size by forming appropriate tensor products with identity matrices before performing the multiplication.}.
A \emph{quantum circuit} corresponds to a sequence of quantum gates or operations that are applied to a certain state.

\begin{figure}[t]
    \centering
    \begin{tikzpicture}
    \begin{yquantgroup} 
    \registers{
    qubit {} q[2];
    }
    \circuit{
init {$\ket{0}$} q[0];
init {$\ket{0}$} q[1];
h q[0];
cnot q[1] | q[0]; 
output {$\ket{\Phi^+}$} (q[0-1]);}

    \circuit{
init {$\ket{0}$} q[0];
init {$\ket{1}$} q[1];
h q[0];
cnot q[1] | q[0]; 
output {$\ket{\Psi^+}$} (q[0-1]);}

    \circuit{
init {$\ket{1}$} q[0];
init {$\ket{0}$} q[1];
h q[0];
cnot q[1] | q[0]; 
output {$\ket{\Phi^-}$} (q[0-1]);}

    \circuit{
init {$\ket{1}$} q[0];
init {$\ket{1}$} q[1];
h q[0];
cnot q[1] | q[0]; 
output {$\ket{\Psi^-}$} (q[0-1]);}

\end{yquantgroup}
    \end{tikzpicture}
    \caption{Generation of Bell states}
    \label{fig:bell}
\end{figure}
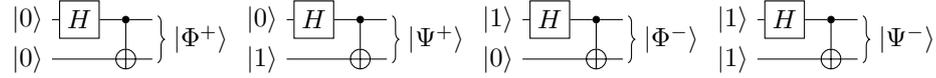

\begin{myexample}\label{ex:bellcircuit}
    Similarly to the classical case, quantum circuits can be illustrated as diagrams where horizontal lines correspond to qubits and gates on the lines correspond to operations acting on the qubits.
    \autoref{fig:bell} illustrates how the four Bell states from \autoref{ex:bellstates} can be generated by starting with any two-qubit computational basis state $\ket{ij}\in\{\ket{00},\ket{01},\ket{10},\ket{11}\}$ and applying a single-qubit Hadamard operation (illustrated as a box labelled $H$) to one of the qubits, followed by a CNOT operation controlled on the same qubit (with $\bullet$ and $\oplus$ indicating the control and the target qubit, respectively).
\end{myexample}

\section{From Encoding Classical Circuits to Quantum Circuits}\label{sec:motiveAndNaive}
In this section, we briefly review how classical circuits and corresponding problems are usually encoded as SAT instances and how these techniques can directly be translated to quantum circuits.
Based on that, we discuss the resulting implications and limitations that form the motivation for this work.

\subsection{Classical Circuits}
In order to obtain a SAT encoding symbolically representing the functionality of a classical circuit, each signal $s$ of the circuit (representing inputs, intermediate signals, and outputs of the circuit) is represented by a corresponding Boolean variable $x^s$.
Then, for each gate $g$ (representing a Boolean operator) of the circuit, a corresponding set of functional constraints relating the input and the output signals of $g$ is introduced.
By further adding constraints to the original SAT formulation of the circuit, e.g., miter structures \cite{brandVerificationLargeSynthesized1993} for verification or formulations for justifying and propagating faults in \emph{Automatic Test Pattern Generation} (ATPG,~\cite{eggersglussImprovedSATbasedATPG2013,DBLP:conf/date/GebregiorgisT19,larrabeeTestPatternGeneration1992,willeSMTbasedStimuliGeneration2009}), a multitude of circuit design tasks can be formulated as SAT instances. Moreover, SAT instances of circuits are frequently used to compute counterexamples, i.e., to show that certain unwanted assignments to the signal variables may occur during the computation of the circuit.

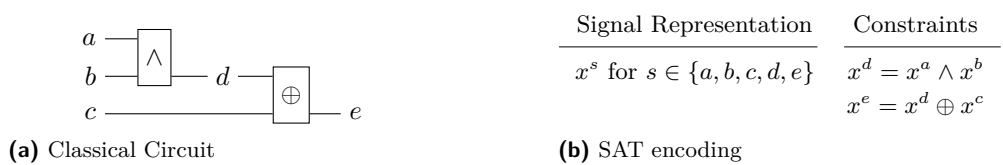
\begin{figure}
    \begin{subfigure}[b]{0.4\textwidth}
    \centering
    \begin{tikzpicture}
        \begin{yquantgroup} 
            \registers{
            qubit {} q[3];
            }
            \circuit{
                init {$a$} q[0];
                init {$b$} q[1];
                init {$c$} q[2];
                hspace {0.3cm} q[0];
                hspace {0.3cm} q[1];
                box {$\land$} (q[0-1]);
                discard q[0];
                hspace {0.3cm} q[1];
                text {$d$} q[1];
                hspace {0.3cm} q[1];
                box {$\oplus$} (q[1-2]); 
                discard q[1];
                hspace {0.3cm} q[2];
                output {$e$} q[2];
            }
        \end{yquantgroup}
    \end{tikzpicture}
    \caption{Classical Circuit \label{fig:ex-classical-sat}}
    \end{subfigure}
    \hspace{0.1\textwidth}
     \begin{subfigure}[b]{0.4\textwidth}
        \begin{tabular}[b]{c  c}
             Signal Representation & Constraints  \\
             \cmidrule(r){1-1}\cmidrule(l){2-2}
             $x^s$ for $s \in \braces{a,b,c,d,e}$ & $x^d = x^a \wedge x^b$ \\
             & $x^e = x^d \oplus x^c$ 
        \end{tabular}
        \caption{SAT encoding}
        \label{fig:ex-classical-sat-encoding}
     \end{subfigure}
     \caption{Classical circuit and corresponding SAT encoding}
\end{figure}

\begin{myexample}\label{ex:classicalSAT}
Consider the classical circuit shown in \autoref{fig:ex-classical-sat}. It consists of two gates, three primary inputs, and a single primary output.
In order to encode this circuit, the signals $a,b,c,d,e$ are represented as Boolean variables $x^a, x^b, x^c, x^d, x^e$, respectively. 
For each logical gate, a corresponding functional constraint representing its functionality is added, as shown in \autoref{fig:ex-classical-sat-encoding}.
\end{myexample}

The main question now is: How can this encoding technique be adapted to the domain of quantum computing?

\subsection{Quantum Circuits}\label{sec:quantum-circuit-encoding}
While each signal in a classical circuit can only assume the value $0$ or $1$, the state of a qubit is generally described as an arbitrary, complex-valued superposition of states,~i.e., $\ket{\varphi} = \alpha_0 \ket{0} + \alpha_1 \ket{1}$. As such, one faces the rather grim perspective of an infinitely-valued logic for even encoding the state of a single qubit.
Consequently, encoding the state vector $\ket{\varphi}$ of a qubit in a quantum circuit (which is the analogue to the signal $s$ in a classical circuit) in a straight-forward fashion is infeasible in general.

However, in practice the number of unique quantum states throughout a quantum computation can be upper bounded since in most cases, the input of a quantum circuit is chosen from a fixed set of states,~e.g., computational basis states $\ket{i} \in \{\ket{0\dots 0},\dots,\ket{1\dots 1}\}$.
In fact, most quantum algorithms assume the initial state to be the all-zero state $\ket{0\dots 0}$\footnote{This is a reasonable assumption, since any initial quantum state can be generated from the all-zero state $\ket{0\dots 0}$ by prepending the actual quantum circuit with a state preparation circuit, e.g., the $\ket{+}$ state can be generated from the $\ket{0}$ state by applying a Hadamard gate.}.
Assuming that a computation may start out in any of $v$ values, the number of different states that are produced by the gates of a quantum circuit $G=g_0,\dots, g_{|G|-1}$ is upper bounded by $2^{|G|} \cdot v$. %
This fact follows from the simple observation that every gate of the circuit can at most double the number of states (by transforming a state $\ket{\varphi}$ to a new state $\ket{\varphi'}$).

Thus, a structural analysis of the complete circuit allows to determine the set of unique quantum states that might occur in a quantum circuit and, accordingly, encode them using a multi-valued logic (in contrast to the binary encoding for classical circuits, as shown in~\autoref{ex:classicalSAT}).
This has already been recognized in~\cite{willeCompactEfficientSAT2013}, showing that, whenever the individual qubits can be considered separately, quantum circuits can be treated similarly to classical circuits.
An example illustrates the idea:

\begin{figure}
    \centering
    
    \begin{tikzpicture}
	\begin{yquant} 
		qubit {} q[2];
		["$a$"]
		box {$\begin{Bmatrix}
				\ket{0} =[0]_{10}\\
				\ket{1} =[1]_{10}
			\end{Bmatrix}$} q[0];
		["$b$"]
		box {$\begin{Bmatrix}
				\ket{0} =[0]_{10}\\
				\ket{1} =[1]_{10}
			\end{Bmatrix}$} q[1];
		hspace {0.5cm} q[0];
		hspace {0.5cm} q[1];
		h q[0];
		h q[1];
		hspace {0.5cm} q[0];
		hspace {0.5cm} q[1];
		["$c$"]
		box {$\begin{Bmatrix}
				\ket{+} =[2]_{10}\\
				\ket{-} =[3]_{10}
			\end{Bmatrix}$} q[0];
		["$d$"]
		box {$\begin{Bmatrix}
				\ket{+} =[2]_{10}\\
				\ket{-} =[3]_{10}
			\end{Bmatrix}$} q[1];
		hspace {3.4cm} q[1];
		hspace {0.5cm} q[0];
		hspace {0.5cm} q[1];
		x q[0];
		hspace {0.5cm} q[0];
		hspace {0.5cm} q[1];
		["$e$"]
		box {$\begin{Bmatrix}
				\; \; \; \,\ket{+} =[2]_{10}\\
				-\ket{-} =[4]_{10}
			\end{Bmatrix}$} q[0] ;
		discard q[0];
		discard q[1];
	\end{yquant}
\end{tikzpicture}
    \caption{Structural analysis of a simple quantum circuit}
    \label{fig:struc-analysis-africon}
\end{figure}
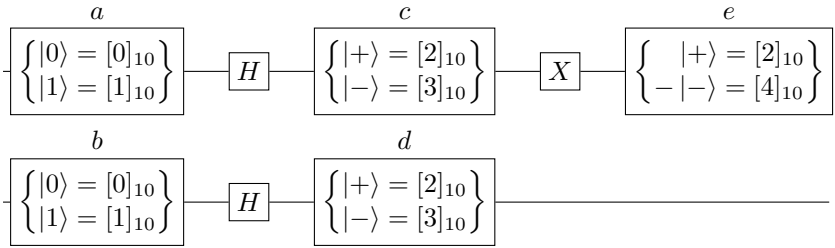

\begin{myexample}\label{ex:structural analysis}
Consider the simple two-qubit quantum circuit composed of three gates shown in \autoref{fig:struc-analysis-africon} and assume that both qubits may either assume the input value $\ket{0}$ or $\ket{1}$.
Since both qubits do not interact over the course of the circuit, the circuit can be split into five signals---labelled $a$ to $e$---in analogy to the classical circuit from \autoref{ex:classicalSAT}.
It is easy to compute that
\begin{equation}
    H \ket{0} =  \ket{+}, \quad H \ket{1} = \ket{-},\quad X \ket{+} = \ket{+}\mbox{, and } X \ket{-} = -\ket{-}.\label{eq:tranformations}
\end{equation}
This simple analysis reveals that there are only five unique states for this particular arrangement: $\ket{0}$, $\ket{1}$, $\ket{+}$, $\ket{-}$, and $-\ket{-}$.
As a consequence, each of the quantum circuit's signals~$s$ can be encoded using $\ceil{log_2 (5)} = 3$ Boolean variables $x_2^s x_1^s x_0^s$.
In a similar fashion as in \autoref{ex:classicalSAT}, the circuit's signals can be related via functional constraints based on \autoref{eq:tranformations}.
\end{myexample}

\subsection{Discussion}\label{sec:naive-enc-discussion}
In general, the feasibility of such an encoding for quantum circuits stands and falls with the feasibility of a structural analysis, i.e., the viability of representing and evolving the state of the quantum system.
In \autoref{ex:structural analysis}, this is feasible since the circuit only involves individual qubits that do not interact.
This kind of analysis extends to the case, where the state of the entire system can always be described as a product state.
A \emph{product state} is a quantum state where the state of the whole system is described as the product of the states of the individual qubits, i.e., $\ket{\varphi} = \ket{\varphi_{n-1}}\otimes\cdots\otimes\ket{\varphi_0}$ with each $\ket{\varphi_i}$ being an arbitrary single qubit state for $i$ from $0$ to $n-1$.
An important subclass of quantum circuits possessing this property are \emph{reversible circuits}\footnote{This assumes that the initial state of the circuit is a classical state, i.e., chosen from the computational basis.}~\cite{frankFutureComputingDepends2017}, which effectively represent bijective Boolean functions and form a dedicated research area on their own.

However, while restricting the potential quantum states to product states allows to efficiently analyze the respective circuits, this restriction prohibits a fundamental quantum mechanical phenomenon to be employed: \emph{entanglement}.
As reviewed in \autoref{sec:background}, two qubits being entangled means that their state \emph{cannot} be considered separately anymore but has to be considered as a whole.
Entanglement is one of the core traits that allows quantum algorithms to surpass classical algorithms for certain applications, e.g., Shor's algorithm for factoring integers is exponentially faster on a quantum computer than on a classical computer~\cite{shorPolynomialtimeAlgorithmsPrime1997}.
Hence, different strategies are needed to replicate the ongoing success of SAT in classical circuit and system design also for quantum circuits.

\section{Generalized Encoding}\label{sec:new_endoding}
In order for the encoding of a quantum circuit to support entanglement, qubits can no longer be considered individually.
As a result, the qubits in a quantum circuit no longer represent individual signals (as in the case of classical circuits or in \autoref{ex:structural analysis}) but are bundled together.
This creates a structure where signals are interleaved with the gates of the circuit and serves to \emph{generalize} the encoding from the previous section. Again, an example illustrates the idea:

\begin{figure}
    \begin{subfigure}[b]{0.75\textwidth}
        \centering
        \usetikzlibrary{quotes}
        \begin{tikzpicture}
		\begin{yquant} 
				qubit {} q[2];
				hspace {0.5cm} q[0];
				hspace {0.5cm} q[1];
				["$u$"]
				box {$\begin{Bmatrix}
						\ket{00} =[ 0]_{10}\\
						\ket{01} =[ 1]_{10}\\
						\ket{10} =[ 2]_{10}\\
						\ket{11} =[ 3]_{10}
					\end{Bmatrix}$} (q[0-1]);
				hspace {0.5cm} q[0];
				hspace {0.5cm} q[1];
				h q[0];
				hspace {0.5cm} q[0];
				hspace {0.5cm} q[1];
				["$v$"]
				box {$\begin{Bmatrix}
						\ket{+0} =[ 4]_{10}\\
						\ket{+1} =[ 5]_{10}\\
						\ket{-0}=[ 6]_{10}\\
						\ket{-1}  =[ 7]_{10}
					\end{Bmatrix}$} (q[0-1]);
				hspace {0.5cm} q[0];
				hspace {0.5cm} q[1];
				cnot q[1] | q[0]; 
				hspace {0.5cm} q[0];
				hspace {0.5cm} q[1];
				["$w$"]
				box {$\begin{Bmatrix}
						 \ket{\Phi^+} =\;[8]_{10}\\
						\ket{\Psi^+} =\;[9]_{10}\\
						\ket{\Phi^-} =[10]_{10}\\
						\ket{\Psi^-} =[11]_{10}
					\end{Bmatrix}$} (q[0-1]) ;
				discard q[0];
				discard q[1];
		\end{yquant}
	\end{tikzpicture}
     \end{subfigure}
        \par\bigskip
     \begin{subfigure}[b]{0.75\textwidth}
         \begin{tabular}{c c c}
             Blocking Constraints & \multicolumn{2}{c}{Functional Constraints} \\
              
              $[\mathbf{x^u}]_2 \leq 3$ &
              $([\mathbf{x^u}]_2 = 0) \Leftrightarrow ([\mathbf{x^v}]_2 =4)$ &
              $([\mathbf{x^v}]_2 = 4) \Leftrightarrow ([\mathbf{x^w}]_2 =8)\;\;$ \\
              $[\mathbf{x^v}]_2 \leq 7$ & 
              $([\mathbf{x^u}]_2 = 1) \Leftrightarrow ([\mathbf{x^v}]_2 =5)$ &
              $([\mathbf{x^v}]_2 = 5) \Leftrightarrow  ([\mathbf{x^w}]_2 =9)\;\;$ \\
              $\;[\mathbf{x^w}]_2 \leq 11$ &
              $([\mathbf{x^u}]_2 = 2) \Leftrightarrow ([\mathbf{x^v}]_2 = 6)$ &
              $([\mathbf{x^v}]_2 = 6) \Leftrightarrow ([\mathbf{x^w}]_2 = 10)$\\
              &
              $([\mathbf{x^u}]_2 = 3) \Leftrightarrow ([\mathbf{x^v}]_2 = 7)$ &
              $([\mathbf{x^v}]_2 = 7) \Leftrightarrow ([\mathbf{x^w}]_2 = 11)$
         \end{tabular}
     \end{subfigure}
     \caption{Structural analysis and resulting encoding for the Bell circuit from \autoref{fig:bell}.\label{fig:bell_structural_analysis}}
\end{figure}

\begin{myexample}\label{ex:naive_bell_encoding}
    Consider again the scenario from \autoref{ex:bellcircuit} and assume we want to perform a structural analysis on the Bell circuit.
    Instead of the two different initial states for each qubit in \autoref{ex:structural analysis} ($\ket{0}$ and $\ket{1}$), the analysis starts off considering four different states of both qubits ($\ket{00}$ $\ket{01}$, $\ket{10}$, and $\ket{11}$).
    Then, the state transitions can be computed in a similar fashion as in \autoref{eq:tranformations}:
    \begin{equation}
        (H\otimes I) \ket{\varphi} = 
        \begin{cases}
            \ket{+}\otimes\ket{0} & \mbox{if } \ket{\varphi}=\ket{00} \\
            \ket{+}\otimes\ket{1} & \mbox{if } \ket{\varphi}=\ket{01} \\
            \ket{-}\otimes\ket{0} & \mbox{if } \ket{\varphi}=\ket{10} \\
            \ket{-}\otimes\ket{1} & \mbox{if } \ket{\varphi}=\ket{11} 
        \end{cases}
        \quad
        \mathit{CNOT}_{1,0} \ket{\varphi} = 
        \begin{cases}
            \ket{\Phi^+} & \mbox{if } \ket{\varphi}=\ket{+}\otimes\ket{0} \\
            \ket{\Psi^+} & \mbox{if } \ket{\varphi}=\ket{+}\otimes\ket{1} \\
            \ket{\Phi^-} & \mbox{if } \ket{\varphi}=\ket{-}\otimes\ket{0} \\
            \ket{\Psi^-} & \mbox{if } \ket{\varphi}=\ket{-}\otimes\ket{1} 
        \end{cases}\label{eq:bell}
    \end{equation}
    Overall, twelve states have to be distinguished in this case.
    Consequently, for each signal~$s$, \mbox{$\lceil\log_2(12)\rceil = 4$} Boolean variables $x_3^s x_2^s x_1^s x_0^s$ are necessary to properly encode the circuit.
    Let~$[\mathbf{x}^s]_2$ denote the interpretation of the variables of the signal~$s$ as a binary number,~i.e., $[\mathbf{x}^s]_2 = \sum_{i=0}^3 2^i x^s_i = [k]_{10}$ for some $k\in\{0, \dots, 2^4-1\}$.
    Then, the valid assignments to each of the signal variables and the functional constraints based on \autoref{eq:bell} can be described by the equations shown on the bottom of \autoref{fig:bell_structural_analysis}.
\end{myexample}

As the previous example shows, a generalization of the encoding from \autoref{sec:motiveAndNaive} indeed allows to incorporate entanglement. In particular, the proposed encoding can be viewed as a two-step procedure:

First, a structural analysis of the circuit is performed to determine the number of unique states that need to be encoded.
This analysis starts off with a set of $v$ possible input states $\{\ket{\varphi_0},\dots,\ket{\varphi_{v-1}}\}$ which are given as an input.
Then, the first gate of the circuit $g_0$ (with corresponding matrix $U_0$) is applied to all states in the initial set of states---yielding a new set of states $\{\ket{\varphi'_0},\dots,\ket{\varphi'_{v-1}}\}$, where $\ket{\varphi'_i} = U_0 \ket{\varphi_i}$ for $i$ from $0$ to $v-1$.
This procedure is then continued for every gate, until the whole circuit has been processed.
Over the course of this process, a separate set $S$ is used to track the unique states.
As already discussed in \autoref{sec:quantum-circuit-encoding}, a maximum of $2^{|G|} \cdot v$ unique states can result from such an analysis.
However, as we will see later in \autoref{sec:experiments}, the number of unique states is typically much lower in practice.

If the first step revealed that there are $|S|$ unique states, $m=\ceil{\log_2(|S|)}$ Boolean variables are needed to encode every signal $s$ in the circuit.
Thus, the state space of a quantum circuit $G$ with $|G|$ gates is encoded using a total of $m\cdot(|G|+1)$ bits, i.e., 
$$ 
\mathbf{x}^s = x_{m-1}^s\dots x_0^s \mbox{ for } s=s_0,\dots,s_{|G|}.
$$
Since $|S|$ might not be a power of two, some of the assignments of the signal variables $\mathbf{x}^s$ do not correspond to valid values.
Let~$[\mathbf{x}^s]_2$ again denote the interpretation of the variables of the signal~$s$ as a binary number,~i.e.,
$$
[\mathbf{x}^s]_2 = \sum_{i=0}^{m-1} 2^i x^s_i = [k]_{10}
$$ 
for some $k\in\{0, \dots, 2^m-1\}$.
Then, so-called \emph{blocking constraints} are introduced to limit the assignments of the signal variables, i.e., 
$$
\forall s\in G\colon\; [\mathbf{x}^s]_2 < |S|.
$$
In addition, the value of the first signal $s_0$ is restricted to one of the $v$ (unique) input states, i.e., 
$$
[\mathbf{x}^{s_0}]_2 < v.
$$
In order to complete the encoding, the circuit's signals have to be related to each other.
Assume that gate~$g_i\in G$ with input signal~$s_i$ and output signal~$s_{i+1}$ maps the $k^\mathit{th}$ unique state~$\ket{\varphi_k}$ to the $l-th$ unique state~$\ket{\varphi_l}$.
Then, the following \emph{functional constraint} is added: 
$$
[\mathbf{x}^{s_i}]_2 = [k]_{10} \iff [\mathbf{x}^{s_{i+1}}]_2 = [l]_{10}.
$$
Eventually, the final encoding is obtained by taking the conjunction over all constraints.

In contrast to the approach described in \autoref{sec:quantum-circuit-encoding}---which directly translates encoding techniques of classical circuits to quantum circuits and, in the process, is limited to only a small fraction of quantum circuits---the generalized encoding proposed above indeed allows to encode the functionality of arbitrary quantum circuits.
However, this comes at a price (\emph{\enquote{there is no free lunch}}\footnote{Adapted from a common version in complexity theory and optimization \cite{DBLP:journals/tec/DolpertM97}.}):
\begin{itemize}
    \item Consider an $n$-qubit quantum circuit and assume that each of the qubits can either be in the $\ket{0}$ or $\ket{1}$ state.
    Then, incorporating entanglement into the encoding, i.e., no longer treating qubits in an isolated fashion, leads to an exponential growth of the potential state space---instead of two states per qubit (as in \autoref{ex:structural analysis}), suddenly $2^n$ states (all of the $n$-qubit computational basis states) need to be considered and encoded.

    \item As reviewed in \autoref{sec:background}, the state of an $n$-qubit quantum system is generally characterized by $2^n$ complex-valued amplitudes.
    Hence, the complexity of representing and manipulating each of the states (i.e., its amplitudes) grows exponentially with respect to the number of qubits---one of the core phenomena that makes simulating quantum circuits on classical computers incredibly hard.
\end{itemize}

This seems like a very grim situation---not only does the encoding result in an exponentially large state space, but also in an exponentially large size of the representation of each individual state. 
Since, as stated earlier, most quantum algorithms actually assume the initial state to be the all-zero state $\ket{0\dots 0}$, only a single initial state needs to be considered in these cases.
However, even then, the exponential size of the states' representation remains a roadblock.
In the following, we show that this dark picture considerably lightens up when not dealing with arbitrary quantum circuits but rather focusing on particular classes of quantum circuits.

\section{Overcoming the Limitations for Certain Classes of Quantum Circuits}\label{sec:cliffordAndBeyond}
As argued in the previous section, an encoding for arbitrary circuits is not feasible in the quantum case due to the sheer complexity of representing quantum states---rendering the structural analysis the dominant source of complexity in the encoding. 
In the following, we demonstrate that an efficient encoding is possible for certain (restricted, yet still important) classes of quantum circuits.
One of the most natural examples, \emph{Clifford} circuits, is discussed first.
Afterwards, we elaborate on other classes of quantum circuits the generalized encoding can be adapted to and briefly discuss merits and drawbacks for each class.

\subsection{Clifford Circuits and the Stabilizer Formalism}\label{sec:clifford}

\emph{Clifford} circuits~\cite{nielsenQuantumComputationQuantum2010}, i.e., quantum circuits generated by the set of gates $\{H, S, \mathit{CNOT}\}$, form one of the most important classes of quantum circuits. This is because
\begin{enumerate}
    \item they describe interesting quantum mechanical phenomena such as entanglement, teleportation, and superdense coding~\cite{nielsenQuantumComputationQuantum2010}, 
    \item they are heavily used in quantum error correcting codes~\cite{calderbank1996good,shor1995scheme,steane1996error}, and
\item according to the Gottesman-Knill Theorem~\cite{gottesman1998heisenberg}, they can be simulated in polynomial time and space on a classical computer using the \emph{stabilizer formalism}.
\end{enumerate}

The main idea of the stabilizer formalism is to represent a quantum state by a set of operators that identify the state uniquely, instead of representing the state by a \mbox{complex-valued} vector of amplitudes. A unitary operator $U$ is said to \emph{stabilize} a state $\ket{\varphi}$ if $U\ket{\varphi} = \ket{\varphi}$, i.e., $\ket{\varphi}$ is an eigenvector of $U$ with eigenvalue $1$.

\begin{myexample}\label{ex:stabilizer}
    Recall the definition of the Pauli matrices $X=\left[\begin{smallmatrix}0&1\\1&0\end{smallmatrix}\right]$ and $Z=\left[\begin{smallmatrix}1&0\\0&-1\end{smallmatrix}\right]$. 
    It holds that $Z\ket{0}=\ket{0}$, i.e., the zero state is stabilized by the Pauli $Z$ operator.
    Furthermore, it holds that $X\ket{+}=\ket{+}$, i.e., the plus state is stabilized by the Pauli $X$ operator.
\end{myexample}

Quantum states that can be produced from the all-zero state $\ket{0\dots 0}$ by applying Clifford gates are frequently called \emph{stabilizer states}.
It can be shown that any $n$-qubit stabilizer state~$\ket{\varphi}$ can be uniquely described by $n$ Pauli tensor products of the form $\pm P_{i,0}P_{i,1}\dots P_{i,n-1}$ with $P_{i,j}\in\{I, X, Y, Z\}$ for $i,j=0,\dots,n-1$.
These $n$-qubit Pauli operators are the generators of the group of stabilizers that stabilize a particular state, i.e., any operator $U$ stabilizing~$\ket{\varphi}$ can be generated from these Pauli operators.
Since two bits are needed for each generator to specify each of the $n$ Pauli matrices and one bit for the sign, $n(2n+1)$ bits are needed in total to uniquely describe the state $\ket{\varphi}$.\vspace{200cm}

\begin{myexample}
    This idea is most conveniently represented via the so-called \emph{stabilizer tableau}~\cite{aaronson2004improved,gottesman1998heisenberg}. 
    For any $n$-qubit state, the tableau consists of $n$ rows for the generators and $2n+1$ columns identifying each generator---which can be grouped as
    $$
      \begin{bNiceArray}{ccc|ccc|c}[]
    x_{0,0} & \Cdots & x_{0,n-1} & z_{0,0} & \Cdots & z_{0,n-1} & r_0 \\
    \Vdots & \Ddots & \Vdots & \Vdots & \Ddots & \Vdots & \Vdots \\
    x_{n-1,0} & \Cdots & x_{n-1,n-1} & z_{n-1,0} & \Cdots & z_{n-1,n-1} & r_{n-1} \\
    \end{bNiceArray}, \mbox{ where }
    \begin{gathered}
        x_{i,j} = \begin{cases}
            1 & P_{i,j} = X \mbox{ or } Y \\ 0 & \mbox{otherwise}
        \end{cases}\\
       z_{i,j} = \begin{cases}
            1 & P_{i,j} = Z \mbox{ or } Y \\ 0 & \mbox{otherwise}
        \end{cases}
    \end{gathered}
    $$
    and $r_i = 1$ if the $i^\mathit{th}$ generator has negative phase for $i,j=0,\dots,n-1$.
    
\end{myexample}

The Gottesman-Knill theorem states that the generators of a stabilizer state can be updated in polynomial time after the application of Clifford gates.
To this end, the following rules for updating the corresponding tableau are used where $\oplus$ denotes a bitwise XOR operation:
\begin{itemize}
    \item $H$ on  $q_j\colon \forall i=0,\dots,n-1\colon r_i = r_i \oplus x_{i,j} z_{i,j}$ and swap $x_{i,j}$ and $ z_{i,j}$, 
    \item $S$ on $q_j\colon \forall i=0,\dots,n-1\colon  r_i = r_i \oplus x_{i,j} z_{i,j}$ and $z_{i,j} = z_{i,j} \oplus x_{i,j}$, and
    \item $\mathit{CNOT}$ with control $q_c$ and target $q_t\colon \forall i=0,\dots,n-1\colon r_i = r_i \oplus x_{i,c}z_{i,t}(x_{i,t}\oplus z_{i,c} \oplus 1)$, $x_{i, t} = x_{i,t} \oplus x_{i,c}$, and $z_{i,c} = z_{i,c} \oplus z_{i,t}$.
\end{itemize}

\begin{myexample}\label{ex:gottesman-knill-bell}
Consider the situation from \autoref{ex:naive_bell_encoding}.
Then, the initial tableau is given by
\[
\begin{bNiceArray}{cc|cc|c}[]
    0 & 0 & 1 & 0 & 0 \\
    0 & 0 & 0 & 1 & 0 \\
    \end{bNiceArray}\mbox{, corresponding to generators } \set{ZI, IZ} \mathrel{\hat{=}} \ket{00}. 
\]
Applying an $H$ gate to $q_1$ leads to the updated tableau
\[
\begin{bNiceArray}{cc|cc|c}[first-row]
      & \textcolor{red}{x_1} & & \textcolor{red}{z_1} & \\
    0 & 0 & 1 & 0 & 0 \\
    0 & \textcolor{red}{1} & 0 & \textcolor{red}{0} & 0 \\
    \end{bNiceArray}\mbox{, corresponding to generators } \set{ZI, IX} \mathrel{\hat{=}} \ket{+0}. 
\]
Finally, applying a $\mathit{CNOT}$ with control $q_1$ and target $q_0$ results in the final tableau
\[
\begin{bNiceArray}{cc|cc|c}[first-row]
    \textcolor{red}{x_t} & \textcolor{red}{x_c} & \textcolor{red}{z_t} & \textcolor{red}{z_c} & \\
    0 & 0 & 1 & \textcolor{red}{1} & 0 \\
    \textcolor{red}{1}& 1 & 0 & 0 & 0 \\
    \end{bNiceArray}\mbox{, corresponding to generators } \set{ZZ, XX} \mathrel{\hat{=}} \ket{\Phi^+}. 
\]
\end{myexample}

Using this stabilizer tableau (rather than exponentially large complex-valued vectors of amplitudes), the individual state representations remain polynomial---since, as demonstrated above, $\mathcal{O}(n^2)$ bits suffice to uniquely describe a stabilizer state.
Hence, instead of directly formalizing the quantum states in the generalized encoding described in \autoref{sec:new_endoding} (as shown in \autoref{ex:naive_bell_encoding}), the respective stabilizer generators can be encoded (in the signal variables~$\mathbf{x}^s$) and related to each other by functional constraints. That is, each of the $k$ unique generators computed in the pre-processing step is encoded using $m = \ceil{log_2(k)}$ variables $\mathbf{x}^s = x^s_{m-1}\dots x^s_0$ such that different assignments to the variables symbolically encode the different generators. Furthermore, respective functional constraints relating the generators to each other are added as described in \autoref{sec:new_endoding}.

As a result, \mbox{SAT-based} solutions (e.g., for equivalence checking) for this central class of quantum circuits with an undoubtedly wide range of applications can be efficiently realized.

\subsection{Beyond Clifford Circuits}\label{sec:beyond}

While Clifford circuits constitute an important class of quantum circuits, it is well known that they are not \emph{universal} for quantum computing~\cite{nielsenQuantumComputationQuantum2010}, i.e., not every quantum computation can be realized using only $H$, $S$, and $\mathit{CNOT}$ (it has even been proclaimed that Clifford circuits are not universal for \emph{classical} computation~\cite{aaronson2004improved}).
Hence, to formulate efficient and scalable SAT encodings for design tasks involving non-Clifford gates, further dedicated classes of quantum circuits need to be considered.
As discussed earlier, the feasibility of the generalized encoding proposed in this work depends on the viability of representing and evolving the state of the quantum system. 
To this end, several data-structures, and methods with the capability of efficiently simulating certain classes of quantum circuits have been proposed. Some of the most important examples are:
\begin{itemize}
    \item \emph{Dense Arrays}: As witnessed in \autoref{sec:motiveAndNaive}, product states, i.e., multi-qubit states where each qubit can be considered in an isolated fashion, can be effectively described by linearly-sized vectors (with respect to the number of qubits).
    \item \emph{Hash Maps}~\cite{jaquesLeveragingStateSparsity2021}: Many state vectors that occur during the simulation of a quantum circuit are sparse due to some structure in the algorithm or even the underlying problem. 
    Thus, such states can be efficiently handled using hash maps instead of storing the dense state vector explicitly.
    \item \emph{Decision Diagrams}~\cite{chin-yungExtendedXQDDRepresentation2011,millerQMDDDecisionDiagram2006,niemannQMDDsEfficientQuantum2016,zulehnerHowEfficientlyHandle2019}: 
    They take the idea of exploiting sparsity one step further by also taking advantage of any redundancies present in the underlying representation.
    To this end, the representation of a quantum state is recursively decomposed and sub-parts only differing by a constant factor are identified with each other---eventually forming a directed, acyclic graph with complex-valued edge weights. Whenever the number of nodes in the decision diagram can be kept low, an efficient representation is obtained.
    \item \emph{Tensor Networks}~\cite{ciracMatrixProductStates2021}:
    Intuitively, tensors can be seen as generalization of matrices and tensor networks are a generalization of quantum circuits that have a similar diagrammatic representation.
    In quantum many-body physics (where, e.g., the collective behavior of interacting particles is studied), it generally holds that particles close to another interact strongly, while particles at a distance hardly interact. 
    This induces a notion of topological locality that naturally motivates modeling such systems as tensor networks.
    Whenever, the dimensions of the tensors and their connections can be kept in check, again, an efficient representation results.
    
    \item \emph{Stabilizer Rank Methods}~\cite{bravyiSimulationQuantumCircuits2019}: 
    The idea of these methods is to decompose a generic quantum circuit into (multiple) stabilizer/Clifford circuits that, as shown in \autoref{sec:clifford}, can be efficiently simulated, and adequately combining the respectively obtained results.
    Their performance scales with the number of non-Clifford gates.
    Hence, as long as this number is low, circuits can be handled efficiently.
\end{itemize}

Overall, if there is an efficient method to conduct a structural analysis of the quantum circuit in question, the generalized encoding proposed in this work can be used to yield a corresponding SAT formulation.
In this fashion, design tasks for a large variety of quantum circuits can be tackled with SAT.\vspace{200cm}

\section{Experimental Evaluation}\label{sec:experiments}
In the following, we provide an empirical analysis on the scalability of the proposed generalized encoding applied to Clifford circuits and demonstrate the feasibility of the pre-processing and construction algorithms for a precise design task (namely \emph{equivalence checking}).
To this end, the generalized encoding proposed in \autoref{sec:new_endoding} has been implemented in C++ on top of the JKQ quantum toolkit~\cite{willeJKQJKUTools2020} using the SAT (SMT) solver Z3~\cite{moura2008z3}.
We used Clifford circuits as a representative class of quantum circuits, since, as described in \autoref{sec:clifford}, this class of circuits is central in quantum computation and can be used to apply the proposed generalized encoding efficiently.
The resulting implementation and evaluation is publicly available at \url{github.com/lucasberent/qsatencoder}.
All experiments were conducted on a machine with an \SI{2.8}{\giga\hertz} Intel i7-1165G7 CPU and \SI{32}{\gibi\byte} RAM running Ubuntu 20.04.

\subsection{Scalability With Respect to Number of Qubits and Gates}

 \begin{figure}[t]
      \centering
     \includegraphics[width=0.78\linewidth]{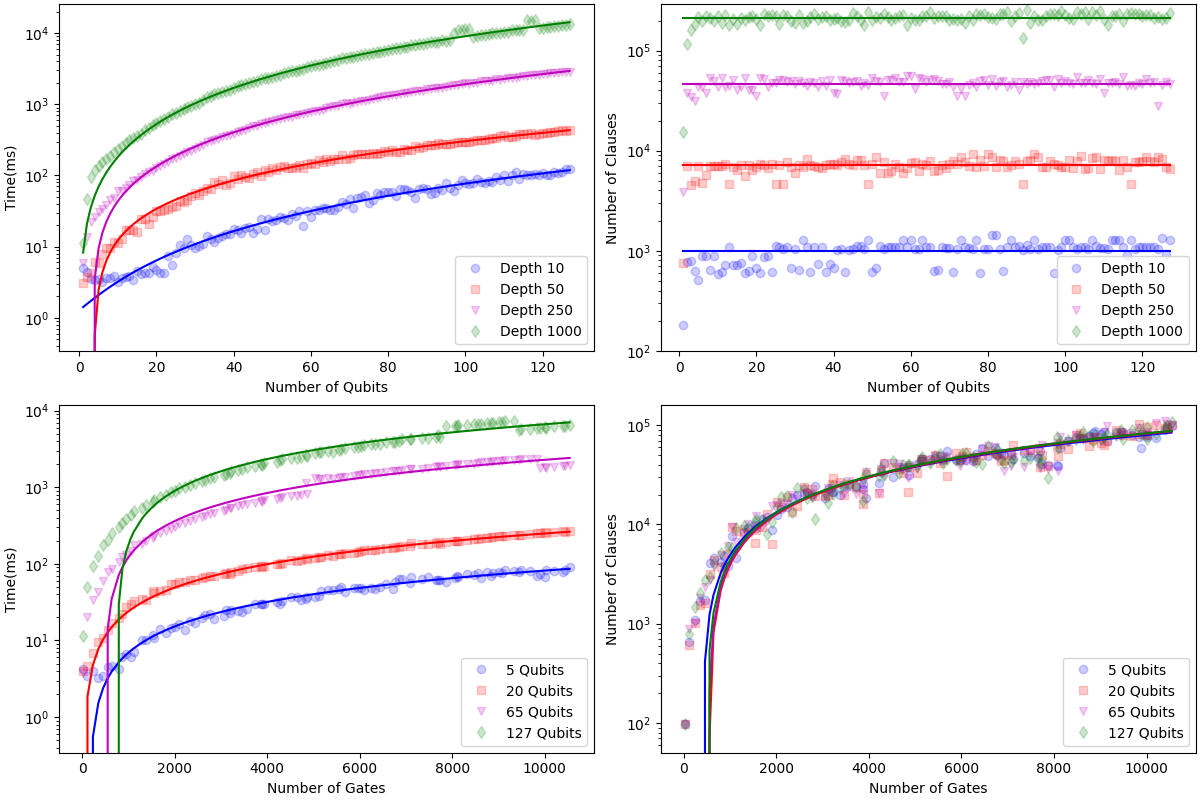} %
     \caption{Scaling of the proposed SAT encoding algorithm applied to Clifford circuits.
    Runtime scales as $\mathcal{O}(n^2)$ and $\mathcal{O}(|G|)$ with $n$ and $|G|$ denoting the number of qubits and gates, respectively.
    The number of clauses is constant in the number of qubits and otherwise scales as $\mathcal{O}(|G|)$.}\vspace*{-3mm}
     \label{fig:scale}
 \end{figure}

In a first series of evaluations, we investigated how fast the encoding for Clifford circuits can be constructed and how many SAT clauses (as reported by Z3) are required with respect to the number of qubits as well as the number of gates. 
To this end, we randomly generated Clifford circuits with growing numbers of qubits and gates. Since the circuits are generated randomly, we sample ten circuits with the same parameters and then plot the mean of the computed values to obtain a representative set of results. 
Without loss of generality, we consider the all-zero state as the single input state. 
The obtained results are depicted in~\autoref{fig:scale}.

The top-left graph shows the time needed (in \si{\milli\second}) to construct a SAT instance using the proposed encoding, given a Clifford circuit. 
This includes the time needed to pre-process the circuit (i.e., to conduct the structural analysis) and to construct the corresponding Z3 instance. 
The graph indicates that the runtime of the proposed algorithm scales quadratically in the number of qubits (which can be attributed to the quadratic size of the stabilizer tableau).

The top-right graph depicts the number of SAT clauses constructed by Z3 with respect to the number of qubits. 
As expected, these functions have constant behaviour in the number of qubits, since, in general, the number of clauses does not depend on that number.
However, the number of qubits remains a dominant factor for the feasibility of the structural analysis.

The bottom two graphs show the runtime and the number of clauses with respect to the number of gates. These results indicate that both, the time needed to construct the encoding and the number of clauses, grow linearly with respect to the number of gates (which can be attributed to the tableau update rules being linear).
Even circuits with \num{10000} gates and \num{127} qubits only roughly take \SI{10}{\second} to encode.

\subsection{Scalability With Respect to Number of Generators}

\begin{figure}[t]
    \centering
    \includegraphics[width=0.95\linewidth]{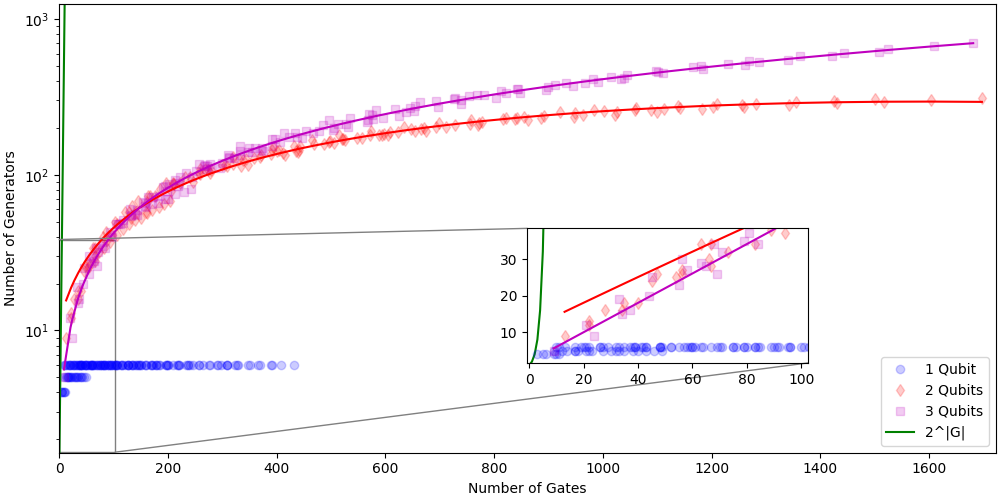}
    \caption{Number of unique generators (states) in Clifford circuits.}\vspace*{-3mm}
    \label{fig:number of generators}
\end{figure} 

As discussed in \autoref{sec:quantum-circuit-encoding}, a maximum of $2^{|G|} \cdot v$ unique states needs to be distinguished in order to encode a particular quantum circuit with $v$ initial input states in the worst-case.
Then, assuming that there are $m$ unique states, a total of $\ceil{\log_2(m)}$ bits are needed to encode each signal of the circuit.
In a second series of evaluations, we investigated how the number of unique states (or, in this case, the number of unique generators) evolves in practice with respect to the number of qubits and the number of gates in Clifford circuits.
As in the previous experiments and without loss of generality, we assume that $v=1$ and that all computations start out in the all-zero state $\ket{0\dots 0}$.

For the experiments, we randomly generated Clifford circuits for one, two, and three qubits with a growing number of gates. Again, we sample ten circuits with the same parameter set and consider the means of the computed values to obtain a representative sample.
The respectively obtained results are shown in \autoref{fig:number of generators}, which also includes a graph for the worst-case complexity $2^{|G|}$.

It can be seen that the actual number of unique generators in Clifford circuits is very small compared to the worst-case behavior, e.g., there are at most $6$ different single-qubit stabilizer states and $360$ two-qubit stabilizer states.
This can be explained by the fact that the size of the stabilizer tableau used to describe particular states inherently limits the number of different states that can occur for a given number of qubits.
As a consequence, at most $\ceil{\log_2(6)}=3$ bits per signal are needed in the single-qubit case, while a maximum of $\ceil{\log_2(360)}=9$ bits is needed for the two-qubit case---independent of the number of gates.

\subsection{Application: Equivalence Checking}

In a final series of evaluations, we demonstrate the applicability of the proposed encoding to the task of verifying the equivalence of two quantum circuits (in particular, Clifford circuits).
To check whether two circuits are equivalent, we construct a \emph{miter} structure~\cite{brandVerificationLargeSynthesized1993}, i.e., the same input is applied to both circuits and the outputs are fed into XOR gates, which are then ORed to produce the single output of the miter.
The task for the SAT solver is to find a variable assignment for which the miter's output evaluates to one.
Whenever this is the case, i.e., the solver returns \emph{satisfiable}, both circuits have been shown to be \mbox{non-equivalent} and the variable assignment provides the counterexample.

For the experiments we again generated random Clifford circuits with a growing number of qubits. 
In a first batch of experiments, we use the same circuit twice and pass the resulting miter structure to the SAT solver (\emph{Equivalent Instances}).
To also test non-equivalent instances, a random single-gate error has been introduced in the second realization of the circuit, i.e., a random gate is removed (\emph{Non-Equivalent Instances}).
In order to detect this non-equivalence, sixteen random input states have been considered for all instances.
Based on~\cite{burgholzerCharacteristicsReversibleCircuits2020}, this amount of input states is almost guaranteed to detect these kinds of errors.

The results are shown in \autoref{tab:ec}.
To this end, the first two columns list the number of qubits $n$ and the number of gates $|G|$.
Then, the runtime for the pre-processing and SAT instance construction $t_\mathit{prep}$, the runtime for the Z3 solver $t_\mathit{sol}$, and the number of conflicts (as reported by Z3) for both, equivalent as well as non-equivalent instances.
These results indicate that even for instance with millions of gates, the respective instances can be constructed in the matter of less than \SI{100}{\second} and solved (i.e., verified) within less than \SI{5}{\second}. 

\begin{table}[tp]
\sisetup{table-text-alignment = right, group-minimum-digits = 4}
\centering
\caption{Equivalence checking for random Clifford circuits with growing number of qubits.\label{tab:ec}}
 \begin{tabular}{@{}*{2}{r}!{\qquad} *{3}{r} !{\qquad}*{3}{r}@{}}%
    \toprule
    \multicolumn{2}{c}{Circuit} & \multicolumn{3}{c}{Equivalent Instances} & \multicolumn{3}{c}{Non-Equivalent Instances} \\
    \cmidrule(r{2em}){1-2}\cmidrule(lr{2em}){3-5}\cmidrule(l){6-8}
     \bfseries $n$ & $\abs{G}$ & $t_{\mathit{prep}}~[\si{\second}]$ & $t_{\mathit{sol}}~[\si{\second}]$ & Confl.  &
     $t_{\mathit{prep}}~[\si{\second}]$ & $t_{\mathit{sol}}~[\si{\second}]$ & Confl. \\
     \midrule
    \begin{filecontents*}{images/ec-out-final.csv}
4,31734,0.815,2.326,28,4,31293,1.238,2.955,8
8,73940,1.204,3.245,36,8,73961,1.489,3.046,2
12,115914,1.576,3.085,22,12,116347,2.07,3.296,0
16,158500,1.905,2.897,15,16,158349,2.576,3.072,0
20,200602,2.671,3.637,33,20,200321,3.275,2.906,7
24,243126,3.118,2.936,15,24,241797,4.172,2.939,2
28,285952,4.093,3.619,37,28,286361,4.784,2.4,5
32,327248,4.989,3.091,30,32,327425,4.922,2.345,2
36,368876,6.492,2.947,16,36,369685,6.499,2.294,0
40,412940,7.79,3.658,33,40,411209,7.641,2.42,2
44,455578,9.022,2.757,18,44,452251,9.015,2.348,2
48,496408,10.605,2.977,15,48,495785,10.59,2.396,2
52,539954,12.351,3.535,34,52,539035,12.183,2.38,6
56,580236,14.039,3.207,20,56,580751,14.197,2.622,0
60,622654,15.816,3.436,32,60,623893,15.778,2.449,1
64,665204,17.822,3.394,30,64,665337,18.028,2.606,0
68,705552,20.283,2.635,13,68,706541,20.487,2.522,0
72,750522,22.599,3.131,15,72,750047,22.699,2.674,0
76,792282,24.998,2.997,15,76,793537,24.854,2.347,9
80,834120,27.329,2.983,15,80,833329,27.361,2.525,0
84,878206,30.022,2.96,17,84,878287,30.153,2.582,0
88,918454,32.606,2.94,18,88,917962,32.768,2.741,0
92,961994,35.808,3.147,15,92,962127,35.639,2.383,8
96,1004410,40.833,3.361,23,96,1003792,40.606,2.819,0
100,1044026,43.828,2.963,15,100,1044725,43.593,2.68,0
104,1085650,47.093,3.106,15,104,1089023,47.129,2.735,0
108,1132082,50.862,2.981,16,108,1131365,51.023,2.337,0
112,1173952,54.299,3.555,36,112,1173675,54.685,2.802,0
116,1214270,78.95,4.55,38,116,1217007,58.83,2.739,0
120,1256578,82.717,4.037,20,120,1253981,62.073,2.677,0
124,1299308,88.647,4.069,20,124,1297929,66.223,2.886,0
\end{filecontents*}
\csvreader[no head,late after line=\\, late after last line=\\\bottomrule, separator=comma]{images/ec-out-final.csv}{}{
		 \csvcoli& 
		 \tablenum[table-format=7]{\csvcolii} & 
		 \tablenum[table-format=2.2,round-precision=2,round-mode=places]{\csvcoliii} & 
		 \tablenum[table-format=1.2,round-precision=2,round-mode=places]{\csvcoliv} & 
		 \tablenum[table-format=2]{\csvcolv} & 
		 \tablenum[table-format=2.2,round-precision=2,round-mode=places]{\csvcolviii} & 
		 \tablenum[table-format=1.2,round-precision=2,round-mode=places]{\csvcolix} & 
         \tablenum[table-format=2]{\csvcolx}}
\end{tabular}\\\vspace{1mm}
{\scriptsize $n$: Number of qubits \hspace*{0.5cm} $\abs{G}$: Number of gates \\ $t_{\mathit{prep}}$: Pre-processing and SAT instance construction time \hspace*{0.5cm} $t_{\mathit{sol}}$: Z3 solving time \\ Confl.: Number of conflicts in Z3 DPLL solving}
\end{table}

\vspace*{-1mm}
\section{Conclusions and Outlook}\label{sec:conclusion}\vspace*{-2mm}
In this work, we investigated the problem of constructing satisfiability encodings for quantum circuits. 
With the goal of establishing efficient SAT formulations for important classes of quantum circuits, we proposed a generalized encoding that can, in principle, be applied to any quantum circuit. 
Since quantum states and circuits require exponentially large representations in general, we identified classes of quantum circuits to which the proposed encoding can be applied in order to obtain efficient satisfiability formulations.

To highlight the practical relevance of the proposed encoding, we considered Clifford circuits, a central class of quantum circuits. 
Our experimental evaluations showed that, for this class of circuits, the proposed encoding can be constructed in an efficient and scalable manner.
The resulting satisfiability formulation can even be generated for large circuit instances with more than one hundred qubits and more than one million quantum gates. 
Modern SAT solvers can then easily solve the instances produced by the proposed encoding technique in a matter of seconds even for numbers of qubits which are currently near the maximum number of qubits in actually available quantum computers.
Therefore, we demonstrated that the proposed formulation can be used to solve highly relevant circuit design tasks for large quantum circuits---as exemplarily showcased by the design task of equivalence checking.

In the future, it will be interesting to apply the generalized encoding to further classes of circuits for which the proposed encoding can, in principle, be constructed efficiently, e.g., those described in \autoref{sec:beyond}, and analyze the expressiveness of these circuit classes and the respective scalability of the proposed encoding.
Furthermore, it will be interesting to start exploring the myriad of design tasks that can now be tackled using SAT techniques.
For all these endeavours, we believe the work summarized in this paper provides a very strong foundation.

\clearpage
\bibliography{lit_header, jku-quantum, refs}
\end{document}